\documentclass[reprint,superscriptaddress,aps,prb,twocolumn]{revtex4-2}
\usepackage{setspace}
\usepackage{amsmath}
\usepackage{graphicx}
\usepackage{subfig}
\usepackage{ragged2e}
\usepackage{amsfonts}
\usepackage{amssymb}
\usepackage{epstopdf} 
\usepackage{xcolor}
\usepackage{verbatim}
\usepackage{soul}
\usepackage{comment}
\usepackage{placeins}
\usepackage{textcomp}
\usepackage{bm}
\usepackage{hyperref}

\DeclareGraphicsExtensions{.pdf,.eps,.png,.jpg,.mps} 

\usepackage{float}

\begin{document}

\title{Building reliable 3D photonic integrated circuits and cavities at the wafer scale}

\author{
Yuhao Huang$^{1}$, 
Yunqi Fu$^{1}$, 
Yu Xia$^{1}$, 
Yuemin Li$^{1}$,\\
Zheng Li$^{1}$, 
Yaoran Huang$^{1}$, 
Zhaoting Geng$^{1}$, 
Mingfei Liu$^{1}$, 
Chao Xiang$^{1,\dagger}$\\
\vspace{0.5em}
$^1$Department of Electrical and Computer Engineering and State Key Laboratory of Optical Quantum Materials, The University of Hong Kong, Hong Kong, China\\
$^\dagger$Corresponding author: cxiang@eee.hku.hk}

\begin{abstract}

Three-dimensional (3D) photonic integrated circuits (PIC) are emerging as an indispensable scheme for high density and multifunctional photonic systems. 
However, the wafer-scale scaling of PICs towards a 3D configuration is constrained by two key factors: (i) the trade-off between inter-layer taper efficiency and footprint, and (ii) wafer-scale uniformity of inter-layer transition loss.
In this work, we introduce etch-back assisted chemical mechanical polishing (E-CMP) to achieve high wafer-scale uniformity of the spacer layer. Moreover, we break the efficiency-footprint trade-off by demonstrating a novel $\kappa$-engineered taper, achieving a reliability metric that is 75\% higher than the traditional linearly tapered structure.
Building on these design and fabrication developments, we enable reliable 3D PICs with typical loss of 0.077 and 0.068 dB/cm on two silicon nitride (SiN) waveguide layers and typical 3D transition loss as low as 6 mdB. 
Furthermore, the low 3D transition loss enables the first class of 3D high-Q optical cavities occupying two distinct device layers, providing new design space for high-Q optical cavities.
The scalable fabrication process and design methodology provide routes for wafer-scale reliable 3D PICs that are promising in a series of applications ranging from photonic interconnects and computing networks to high-density photonic sensors and nonlinear photonics.

\end{abstract}

\maketitle

\section{Introduction}

The rapid advancement of photonic integrated circuits (PIC) is projected to meet the exponentially growing demands of modern data communication systems and future hyper-scale computing applications. However, the footprint miniaturization of PICs is fundamentally restricted by the light-confining characteristics of optical waveguide (WG), which, for planar PICs, impedes device density enhancement and complicates WG routing for large scale systems~\cite{zhang2020scal, zhu2024low, Xia2025ThreeD}. Moreover, emerging potential functionalities of PICs are being progressively unlocked by hybrid/heterogeneous integrations that are based on multilayer, vertically stacked architectures~\cite{kaur2021hybrid}. These features are driving the evolution of PICs toward three-dimensional (3D) integration frameworks with enhanced density~\cite{jones2013ultra, sacher2018mono, Margalit2021Per} and comprehensive functionalities~\cite{tran2022ext, churaev2023hg, xiang20233d}, promising to achieve breakthrough performance in next-generation photonic systems.

\begin{figure*}[ht]
	\centering
	\includegraphics[width = 1 \textwidth]{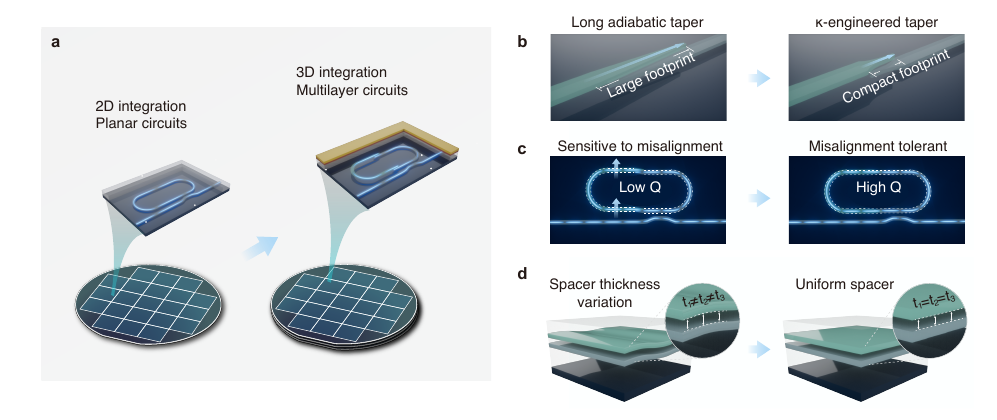}
	\caption{
    Challenges and solutions for wafer‑scale 3D PICs.
    (a) Schematic illustration depicting the transition from 2D to 3D PIC architectures.
    (b) The proposed $\kappa$-engineered taper overcomes the inherent trade-off between inter-layer taper efficiency and device footprint, enabling compact and high-performance 3D PICs.
    (c) Lateral misalignment during fabrication induces significant inter-layer transition losses in conventional adiabatic tapers, a limitation that is effectively mitigated by the $\kappa$-engineered taper, thereby enhancing optical performance.
    (d) Variations in vertical spacer layer thickness compromise the yield and reliability of 3D PICs, which are resolved through fabrication process development to ensure uniformity and reliability.
    }
	\label{fig1}
\end{figure*}

For 3D PICs to perform optimally, every layer must operate in a fully integrated manner, exhibiting high uniformity and minimal loss similar to what is achieved within a single-layer platform.
Without such universal reliability, the practical usability of 3D PICs will be severely constrained, since unpredictable inter-layer transition efficiency would remain a major obstacle in 3D PICs designs.
However, the revolution to reliable 3D PICs, especially at the wafer-scale, comes with significant design and fabrication challenges, as shown in Fig.\ref{fig1}.
First, components for layer-to-layer transition still occupy significant chip area. Despite sufficient optical bandwidth and fabrication simplicity over vertical via~\cite{zhang2018high} and grating coupler~\cite{kang2014amorphous, sodagar2014high, wan2016grating, xu2019sin}, adiabatic tapers for layer-to-layer transition raise a trade-off as high efficiency requires significant long taper length, which restricts the freedom of layout design when a layer-to-layer transition occurs.
Second, non-uniform behavior of taper transition loss across the wafer increases the loss budget in optical links, restricting the maximum number of achievable cross-layer connections and limiting the 3D integration density.
In practical taper components, transition loss mainly stems from the lateral and vertical offsets from the design values~\cite{shang2015low}. The former is governed by the alignment precision of lithographic or bonding systems, while the latter depends on the uniformity of spacer SiO$_2$ deposition and planarization. Moreover, the origins of vertical deviation often lead to a random distribution, resulting in varying taper transition loss even within the same die. Such variability disqualifies the control of wafer-scale uniformity and constrains the potential for mass production of 3D PICs. 
Consequently, to attain reliable 3D PICs, we need efficient taper structures with high misalignment tolerance for rapid layer-to-layer transition, as well as a fabrication process enabling uniform cross-wafer control of vertical spacer layers to support high production yield.

Several studies have demonstrated the feasibility of leveraging inter-layer transitions to realize 3D PICs, but important limitations remain. For instance, tri‑layer SiN-SiN-Si platform achieved low crosstalk at the expense of excessively long structures (70+120 $\mu$m) and strong dependence on layer uniformity, with misalignment studies restricted to narrow ranges ($\pm60$ nm)~\cite{sacher2017tri}. Similarly, SiN–AlN integration operating over a multi-octave range employed adiabatic tapers with coupling lengths exceeding a demanding value of 1000 $\mu$m, and misalignment tolerance was not reported~\cite{zhang20253d}. In addition, high‑speed 3D heterogeneous LiTaO$_3$ platform demonstrated $>$70 GHz operation yet suffered from large inter-layer transition loss (0.6 dB) and significant sensitivity to lateral misalignment when using patterned LiTaO$_3$ ridge waveguides~\cite{niels2026high}. Across these approaches, while adiabatic couplers remain the most promising solution for 3D PICs, current demonstrations often equate reliability with the sparse set of experimental data. True reliability, however, is not a single low‑loss metric but the combination of design robustness, process tolerance, and statistical consistency across large-scale production.

In this work, we demonstrate a reliable 3D PIC on a dual-layer 400-nm-thick SiN platform with wafer-scale uniformity in both intra- and inter-layer loss performance. 
The uniformity of spacer layer across the wafer-scale is guaranteed through etch-back assisted chemical mechanical polishing (E-CMP) and stress release trenches (SRT), which contributes to significant uniformity as the Coefficient of Variation (CV) of WG propagation loss in Layer 1 (L1) and Layer 2 (L2), and spacer thickness are 11.92\%, 8.99\%, and 2.06\%, respectively, and the corresponding Kurtosis ($\beta_2$, a measure of distribution shape) are 2.97, 2.04, and 2.01, respectively. 
We further propose a novel $\kappa$-engineered taper ($\kappa$-taper) to break the efficiency-footprint trade-off, and it outperforms the traditional linearly tapered structure (L-taper) by 75\% in reliability metrics evaluated by inter-layer transition efficiency and lateral misalignment tolerance.
Eventually, building on the low-loss multilayer SiN platform and the reliable $\kappa$-taper, we present a reliable 3D PIC that achieves an inter-layer transition loss as low as 5.56 mdB/coupler, with wafer-scale analysis yielding a CV of 9.69\% and a $\beta_2$ of 2.78.
Furthermore, these exceptionally low‑loss inter-layer transitions enable a new class of high‑Q, 3D‑integrated optical cavities that introduce additional design freedom for advanced resonator architectures.
The proposed method effectively mitigates the reliability challenges in 3D PICs fabrication, providing a standardized manufacturing route capable of enabling high-density and multifunctional 3D photonic chips.

\section{Results}

\subsection*{Addressing Stress-Induced Non-Uniformity in 3D SiN PIC}

\begin{figure*}[ht]
	\centering
	\includegraphics[width = 1 \textwidth]{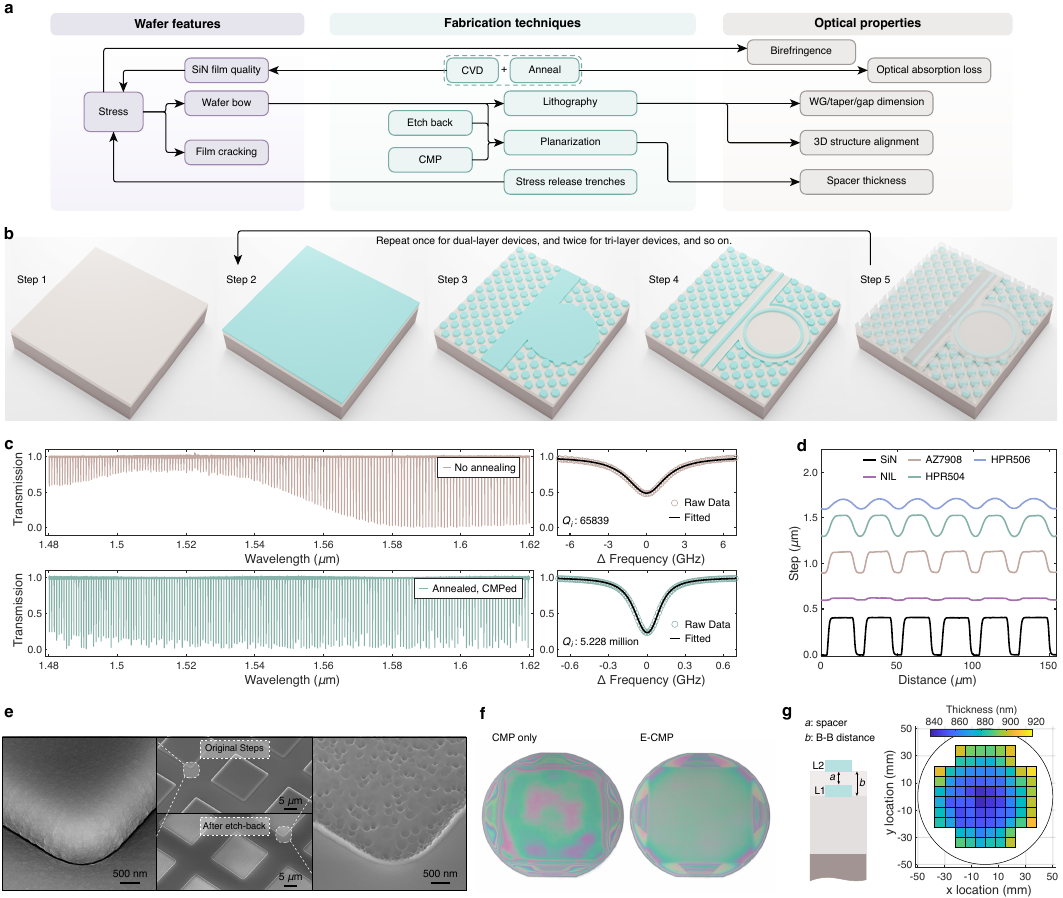}
	\caption{
    Stress challenges and corresponding solutions for 3D SiN devices. 
    (a) The relationship among wafer defects, fabrication techniques, and the optical properties; 
    (b) The fabrication process consists of a standardized 5-step sequence. For multilayer chips, the same sequence is iterated. 
    (c) Left: Comparison of SiN WG with and without annealing and CMP with wavelengths ranging from 1480 nm to 1620 nm. Right: The corresponding resonance dips around 1550 nm. 
    (d) Steps coverage abilities of different photo resists. 
    (e) The profile of 400-nm-high steps before and after etch-back. 
    (f) Comparison of the wafer images using only CMP and E-CMP.
    (g) Left: The diagram showing the definition of spacer and B-B distance. Right: Measured distribution of the B-B distance for the 3D PIC demonstrated in this work, obtained using SRT and E-CMP.
    }
	\label{fig2}
\end{figure*}

Owing to its broad optical transparency and low refractive index contrast with SiO$_2$, SiN is widely employed to fabricate low-loss WG~\cite{bauters2011ultra, jin2021hertz, ji2022compact, ji2024efficient}, and it also demonstrates significant potential for 3D PICs applications.
However, to achieve 3D SiN PICs at the wafer-scale, tensile stress in SiN film resulting from CVD (chemical vapor deposition of film) and high-temperature annealing (film purification) poses a major challenge (Fig.\ref{fig2}a). Generally, wafer bow can impact depth-of-focus variations during stepper lithography, which notably decreases the yield of precise fabrication of key photonic structures, including gaps between WG and tapered structures with small tips. 
For 3D PICs, severe wafer bow results in the non-uniform behavior of taper transition loss. 
For lateral offsets, wafer bow can affect lithographic misalignment across dies, which results in overlay misalignment that is uncorrectable by a global shift of the lithography exposure field (see details in Supplementary Information, SI Section I). 
For vertical offsets, on the other hand, wafer bow poses difficulties in chemical mechanical polishing (CMP) and leads to non-uniform formation of the vertical spacer between the two WG layers, thereby introducing vertical deviations.
As stress further increases, SiN film can even crack and lead to complete device failure and production waste.

Aiming at low tensile stress, plasma-enhanced chemical vapor deposition (PECVD) is used to prepare SiN film since it features intrinsic low stress and provides convenient stress management and fabrication repeatability, especially for 3D PICs with multi-step overlay. Detailed comparison of SiN deposition method is discussed in Methods. After deposition, the SiN top surface smoothness is improved by a mild CMP, reducing the RMS roughness from 1.05 nm to 0.29 nm (see FigS5b in SI), which helps reduce the scattering loss. To further lower the optical propagation loss, high-temperature annealing is indispensable, but it also leads to considerable stress. Therefore, we etch SRT prior to each annealing cycle to divide the continuous SiN film into localized segments since the macroscopic impact of stress can be significantly diminished when the SiN film is first scaled down to areas of tens of microns before annealing. The fabrication process flow is shown in Fig.\ref{fig2}b and described in Methods. With SRT confining the annealing-caused stress to discrete regions, crack propagation from the no-SRT zone to the SRT-treated area is successfully prevented, and the wafer bow is well controlled (see details in SI Section II). For single‑mode 3D photonic integration, SiN WG of 400 nm thickness and 900 nm width is fabricated, and the loss is characterized with ring resonators (overall spectra shown in Fig.\ref{fig2}c). PECVD SiN resonators without annealing show an intrinsic quality factor ($Q_i$) of 0.066 million near 1550 nm, while after the annealing and CMP process, $Q_i$ is improved by two orders of magnitude to around 5.2 million.

The SiO$_2$ planarization procedure attributes to another origin of spacer non-uniformity. Despite the optimized CMP recipe, SiO$_2$ spacer uniformity remains limited by step height, pattern density, etc., worsening with prolonged CMP for desired flatness and smoothness. Therefore, we introduce E-CMP process, using etch-back for step height reduction and short-time CMP for surface smoothness enhancement, to achieve high wafer-scale uniformity of the spacer layer. Fig.\ref{fig2}d compares the step coverage abilities of different photo resists, indicating that the optimal solution among the tested samples is nanoimprint lithography (NIL) resist, with 600 nm thickness and achieving nearly 94\% height reduction, from 400 nm to 20–25 nm (detailed development of E-CMP is introduced in SI Section IV). After the SiO$_2$ steps are covered by NIL resist, they are etched at the same rate by an inductively coupled plasma (ICP) dry etcher, as shown in Fig.\ref{fig2}e. The residual topography is removed by a short‑time CMP, and an RMS roughness of 0.2 nm is achieved. Fig.\ref{fig2}f shows that the wafer processed by only CMP exhibits nonuniform white light interference patterns due to large SiO$_2$ spacer thickness variations, while the wafer processed by etch-back-assisted CMP displays a notable uniformity improvement. To accurately measure the SiO$_2$ thickness, it is essential to minimize the number of films composed of different materials during fitting the film thickness. Therefore, the end-point detection of CMP should measure the bottom-to-bottom distance (B-B distance) between L1 and L2, which is schematically defined in Fig.\ref{fig2}g. For the designed 450-nm-thick spacer layer on a 400-nm-thick SiN WG, B-B distance is 850 nm. The measured wafer-scale B-B distance after E-CMP is summarized in Fig.\ref{fig2}g and Table \ref{T1}, which proves to be markedly uniform across the 4-inch wafer with CV and $\beta_2$ being 2.06\% and 2.01, respectively.

So far, we have developed fabrication techniques for 3D SiN that exhibit intrinsically low tensile stress, low optical WG propagation loss, and precisely controllable inter-layer spacer thickness, which together mitigate the non-uniformity in 3D PIC. It needs to be noted that the approach is also valid for fabricating thick SiN WG ($>$800 nm) using single step PECVD deposition and a subtractive process (see SI Section III for more details), potentially offering a standardized approach to stress management for SiN devices from thin to thick dimensions and from single to multilayer to allow performance control such as dispersion engineering.

\subsection*{Rapid and robust inter-layer transition enabled by $\kappa$-engineered taper}

\begin{figure*}[ht]
	\centering
	\includegraphics[width = 1 \textwidth]{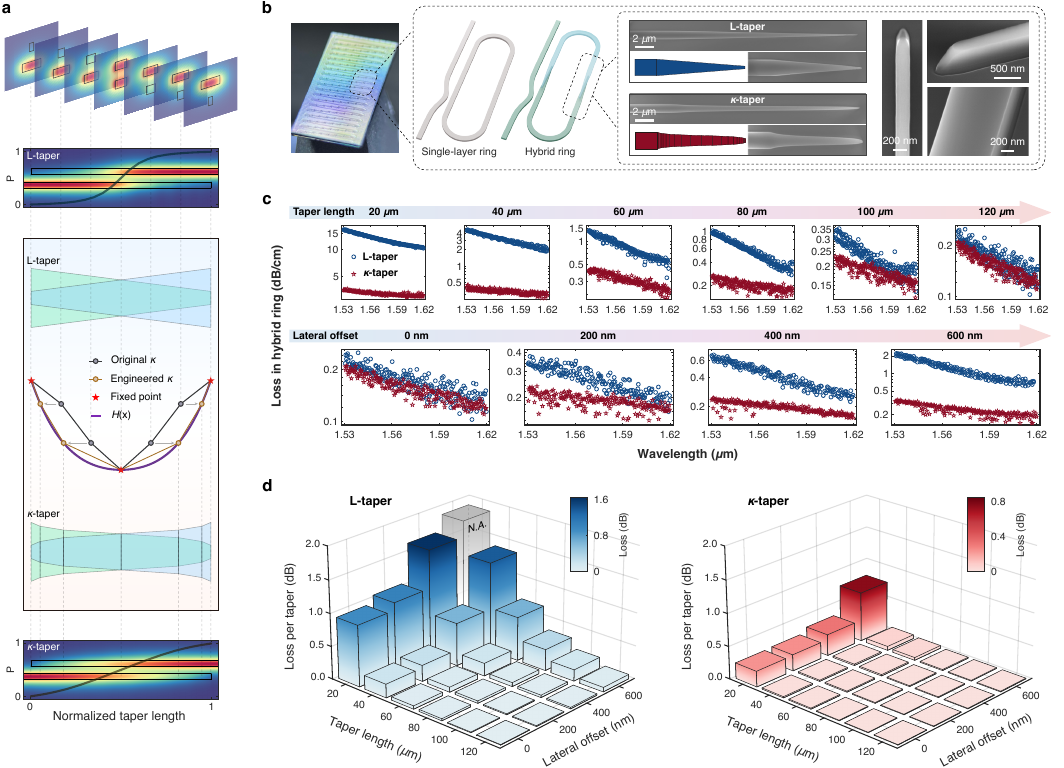}
	\caption{
    $\kappa$-taper for rapid and robust inter-layer transition. 
    (a) Conceptual illustration of the design principle of $\kappa$-taper, showing the remapping process of $\kappa$ and thereby reshaping the taper geometry.
    (b) Photographs showing the fabricated 3D SiN chip. Left: ring resonators for loss measurement. Middle: SEM of L-taper and $\kappa$-taper. For the images without scale bar, the horizontal scale is compressed by a factor of 4 to facilitate visualization. Right: SEM details the taper tip and WG sidewall. 
    (c) Propagation loss (dB/cm) in hybrid ring vs. wavelength for L-taper and $\kappa$-taper with varying taper length and lateral offset.
    (d) Taper loss extracted from ring resonances for comparison of L-taper and $\kappa$-taper under different taper lengths and lateral offsets. In the group combining a 600 nm offset with a 20-$\mu$m-long device, L-taper introduces high transition loss and leads to the failure of resonance fitting, resulting in a not applicable loss value.
    }
	\label{fig3}
\end{figure*}

To address the efficiency-footprint trade-off and the sensitivity to lateral/vertical offset in traditional adiabatic tapers, we propose $\kappa$-engineered taper ($\kappa$-taper) for rapid and robust inter-layer transition. The $\kappa$-engineering initiates with the dual-layer linearly tapered coupler (L-taper), with the outline and side view of the power transition illustrated in Fig.\ref{fig3}a. The coupling coefficient $\kappa$ of L-taper decreases to the minimum value in the middle of the structure as the two tapered waveguides share the identical width. This location is the phase-matched point where the power abruptly transfers from the lower taper to the upper one, as indicated by the dark line alongside the side view of the power transition of L-taper. This feature leads to two significant shortcomings: (i) L-taper with weak and sharp $\kappa$ around the phase-matched point requires an extremely long taper length to achieve high power transmission. (ii) Abrupt power transition around the phase-matched point results in low fabrication tolerance as tiny lateral and vertical offsets will bring large perturbation and greatly reduce transmission. As a result, when $\kappa$ can smoothly vary around the phase-matched point, the above-mentioned drawbacks can be mitigated. 

To flatten the curve of $\kappa$, Hermite interpolation is employed to create a gently varying curve $H(x)$ at the desired points and derivatives, and then the $\kappa$, together with the taper segments, is remapped. The optimization of derivatives and the number of segments are discussed in Methods and SI Section V. The $\kappa$-taper demonstrates two significant advantages that precisely address the two weaknesses of L-taper: (i) The value of $\kappa$ near the phase-matched condition takes up a large area of the overall taper. When the phase-matched `point' is expanded to a large `area', as shown in the side view mode profile in Fig.\ref{fig3}a, the entire structure is more efficient at facilitating the power transfer, contributing to a more compact footprint than the linear taper. (ii) $\kappa$-taper embodies gentle power transition, as indicated by the dark line alongside the side-view of the power transition of $\kappa$-taper, and therefore it suffers less from device non-uniformity (lateral and vertical deviation) and provides a high fabrication robustness.

While existing research on taper length reduction has mainly focused on theoretical calculations~\cite{liang2021fully, taras2021shortcuts, yao2022bridging, van2024methodical}, our approach combines theoretical simulations with systematic experimental validation across a range of taper lengths and offsets, thereby providing robust evidence for the reliability of the proposed $\kappa$-taper. We fabricate a dual-layer SiN chip using the above-introduced fabrication techniques to study the taper performance. Fig.\ref{fig3}b displays the essential devices for loss measurement: single-layer ring resonators on L1 and L2 for WG loss measurement, and ring resonators across both layers (called hybrid resonators) for inter-layer transition loss measurement. Assume that the propagation losses of resonators (on L1/L2 and hybrid type) are $\alpha_{1,2,hybrid}$, the resonator circumference is $L_{ring}$, and taper length is $L_{taper}$. Then, the transition loss for inter-layer coupler ($\alpha_{taper}$, dB/coupler) can be written as:

\begin{equation}
    \alpha_{taper} = \frac{1}{2}\left[\alpha_{hybrid}L_{ring}-\frac{1}{2}(\alpha_{1}+\alpha_{2})(L_{ring}-2L_{taper})\right]
    \label{e1}
\end{equation}

Each resonator has a 1499 $\mu$m circumference, corresponding to a free spectral range of 100 GHz. The SEM images in the middle panel of Fig.\ref{fig3}b display the fabricated L-taper and $\kappa$-taper, with SEM images in the middle row compressed by 4 times in the horizontal direction for visual comparison of the designed taper geometry and actual devices. SEM images on the right demonstrate the fabricated taper with 200-nm-wide tip and WG with smooth sidewalls. Two tapered structures are intentionally offset in the layout design to emulate potential lateral misalignment. The thickness of SiO$_2$ layer across the 4-inch wafer is shown in Fig.\ref{fig2}g.

To minimize the influence of variations in the vertical spacer, chips positioned in the middle of the wafer (die \#6 and die \#11, see Fig.\ref{fig4}b for die number) are chosen for further analysis. Fig.\ref{fig3}c shows that $\kappa$-taper consistently exhibits superior performance over L‑taper across the investigated parameter space of taper length (20–120 $\mu$m) and lateral offset (0–600 nm). For different taper lengths, $\kappa$-taper maintains remarkably high transmission efficiency even as the taper length is substantially reduced. For instance,  $\kappa$-taper with a length of 40 $\mu$m exhibits better loss performance compared to the 80-$\mu$m-long L-taper. This feature enables efficient energy transfer within a compact footprint, effectively breaking the trade-off between efficiency and device footprint. Similarly, under different lateral offsets, the $\kappa$-taper maintains comparatively low loss, while the L‑taper shows rapid degradation in transition efficiency, underscoring its great sensitivity to misalignment. On the other hand, the spectral loss distribution of the $\kappa$-taper in Fig.\ref{fig3}c is flatter than that of the L-taper, suggesting lower wavelength dependence that enables the $\kappa$-taper to exhibit a broad optical bandwidth. Fig.\ref{fig3}d summarizes the taper loss data from all 24 experimental groups by extracting the loss from resonance dips and reveals a significant trend that $\kappa$-taper can sustain its exceptionally low transmission loss even when subjected to short taper lengths and severe lateral deviations. For instance, in the experimental group that combines a 600 nm offset with a 20-$\mu$m-long device, L-taper introduces such high transition loss that the transmission spectrum of the hybrid ring resonators can hardly be fitted, whereas $\kappa$-taper maintained a low loss of 0.76 dB/coupler. A dimensionless reliability metrics ($R$) evaluated by transition efficiency ($T$) at specific lengths ($L$) and lateral offsets ($M$) is also proposed to characterize the taper performance:

\begin{equation}
    R = T\exp{\left(\frac{120}{L}\right)}\exp{\left({\frac{M}{600}}\right)}
    \label{e2}
\end{equation}

Considering all experimental groups in Fig.\ref{fig3}d, the mean values of $R$ for $\kappa$-taper and L-taper are 117.9 and 67.3, respectively, indicating that $\kappa$-taper outperforms the L-taper by 75\%. The proposed $\kappa$-taper successfully breaks the efficiency-footprint trade-off and features superior tolerance to misalignment. The exceptionally low transmission loss allows the lightwave to transfer freely from layer to layer, providing a clear pathway toward higher‑yield and uniform wafer‑scale 3D PICs. 

\subsection*{3D PICs and cavities with wafer-scale uniformity}

Building upon the fabrication developments and the benefits of $\kappa$-tapers, we enable wafer-scale reliable 3D PICs and novel 3D high-Q optical cavities.

\begin{figure*}[]
	\centering
	\includegraphics[width = 1 \textwidth]{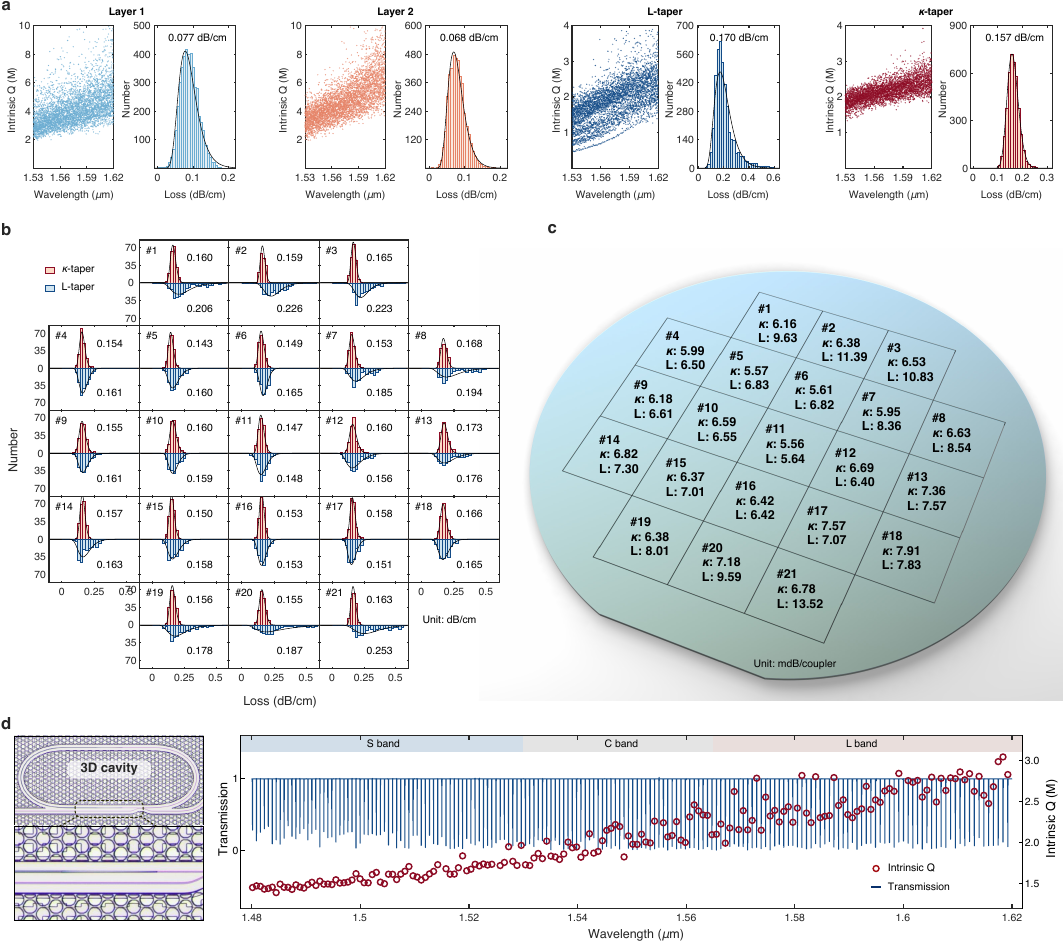}
	\caption{
    Reliable 3D PICs and cavities at the wafer-scale. 
    (a) $Q_i$ and the corresponding loss histograms of the resonators in L1, L2, and hybrid resonators (consisting of 120-$\mu$m-long L-taper and 120-$\mu$m-long $\kappa$-taper) from 21 dies across the 4-inch wafer, measured from 4385, 4430, 4170, and 4074 resonant dips, respectively.
    (b) Comparative histograms of ring propagation loss (dB/cm) for L-taper and $\kappa$-taper across a 4-inch wafer map.
    (c) Wafer-level distribution of transition loss (mdB/coupler) for L-taper and $\kappa$-taper across 21 dies.
    (d) Left: microscopic image showing a 3D cavity that occupies two SiN layers. Right: the spectrum and $Q_i$ of this single cavity that features high-Q across the telecom S, C, and L bands.
    }
	\label{fig4}
\end{figure*}

Figs.\ref{fig4}a demonstrate the overall wafer level statistical analysis of the resonators considering all 21 different dies with wavelengths ranging from 1530 nm to 1620 nm. For devices in L1, the scatter diagram in Fig.\ref{fig4}a reveals the wavelength dependence of $Q_i$ extracted from 4385 resonances from 1530 nm to 1620 nm. Then, the WG propagation loss is calculated from $Q_i$ and presented in the histogram, with the most probable value in the histogram being 0.077 dB/cm. For resonators in L2, the most probable propagation loss is 0.068 dB/cm by valuing 4430 resonances. Similarly, for hybrid resonators spanning both layers, L-taper-based resonators exhibit a propagation loss of 0.170 dB/cm (valuing 4170 resonances), whereas $\kappa$-taper-based resonators achieve a lower loss of 0.157 dB/cm (valuing 4074 resonances). Although the most probable loss values for both resonators based on different tapers are similar, the histogram data show a narrower loss distribution for $\kappa$-taper-based resonators, indicating superior performance consistency in $\kappa$-taper at the wafer-scale. From data in Fig.\ref{fig4}a, the wafer level transition loss can be calculated using Equation (\ref{e1}), yielding 8.18 mdB/coupler for L-taper and 7.20 mdB/coupler for $\kappa$-taper.

\begin{table}[ht]
\centering
\caption{Statistical analysis for D2D variation}
\label{T1}
\begin{tabular}{c|cccc}
\hline
                              &  Mean              & CV          & $\beta_2$         \\ \hline
\textbf{B-B distance}               & 872 nm             & 2.06\%      & 2.01              \\
\textbf{L1 WG loss}           & 0.0894 dB/cm       & 11.92\%     & 2.98              \\
\textbf{L2 WG loss}           & 0.0784 dB/cm       & 8.99\%      & 2.05              \\ 
\textbf{L-taper loss}         & 8.02 mdB/coupler   & 24.58\%     & 4.11              \\ 
\textbf{$\kappa$-taper loss}  & 6.51 mdB/coupler   & 9.69\%      & 2.78              \\ \hline
\end{tabular}
\end{table}

On the other hand, Fig\ref{fig4}b and Fig\ref{fig4}c demonstrate the die level statistical analysis of the hybrid resonators from 21 different dies. The die level statistical analysis of L1/L2 WG loss is discussed in SI Section VI. Fig\ref{fig4}b displays the histograms of propagation loss in hybrid resonators for 21 dies, and each histogram comprises an upper part for $\kappa$-taper-based resonators and a lower part for L-taper-based resonators, with the most probable loss value denoted. In particular, Fig\ref{fig4}b reveals the unsatisfactory performance of L-taper located on the wafer outer region. This phenomenon results from the deviation of the spacer thickness on the wafer outer region, as shown in Fig.\ref{fig2}g, which makes the L-taper at the wafer edge (die \#1, \#2, \#3, \#8, \#13, \#18, \#19, \#20, and \#21) exhibit histograms with obvious tails toward higher loss regions, even though their modal values approximately align with those in other dies. On the contrary, the propagation loss in $\kappa$-taper-based hybrid resonators features a more centralized distribution than their L-taper counterpart for every die. This phenomenon further confirms that $\kappa$-taper exhibits strong resistance to vertical spacer deviation, whereas L-taper proves to be markedly sensitive to such variation. The wafer-scale perspective of transition loss for inter-layer coupler can be obtained by applying Equation (\ref{e1}) to each die, and the results are shown in Fig.\ref{fig4}c. The inter-layer transition loss as low as 5.56 mdB/coupler is achieved by $\kappa$-taper, and the excellent uniformity is also demonstrated across the entire wafer.

Further analysis for die-to-die (D2D) variation is evaluated by CV and $\beta_2$, with the results displayed in Table \ref{T1}. For L1/L2 WG loss, CV are 11.92\% and 8.99\%, respectively, indicating low-loss features with wafer-scale uniformity, whereas a CV of 24.58\% for the L-taper transition loss shows moderate variability, primarily due to its significant sensitivity to spacer variation at the wafer edge. In contrast, a CV of 9.69\% for the $\kappa$-taper transition loss signifies a wafer-scale consistency, even though the spacer thickness at the wafer edge varies from design. On the other hand, $\beta_2$ of B-B distance, L1/L2 WG loss, and $\kappa$-taper loss all fall below 3, characterized by a flatter peak and thinner tails than the normal distribution, which implies a lower likelihood of extreme loss value.

The exceptionally low-loss inter-layer transitions allow a single high-Q cavity to occupy two distinct layers. Fig.\ref{fig4}d displays 3D $\kappa$-taper-based hybrid ring cavity along with its measured transmission spectrum and $Q_i$. As the transition loss reduces to a level of mdB per coupler, light propagation across multiple layers becomes very close to in-plane transmission, and the hybrid ring can thus operate in a fully integrated manner. Therefore, the intrinsic loss within the cavity is governed by the waveguide loss, and it can feature high-Q properties across the telecom S, C, and L bands. This enables distinct layers, more importantly, multiple materials, to be potentially integrated not only through physical bonding but also via seamless optical transitioning, effectively creating a continuous and low-loss photonic pathway across heterogeneous platforms. Therefore, building 3D cavities are promising to leverage the optical properties in a new dimension, providing a novel approach to unlock intriguing applications ranging from interconnects and computing networks, to high-density photonic sensors and nonlinear photonics.

\section{Discussion}
We have presented a reliable 3D photonic integrated circuits and novel 3D cavity at the wafer-scale through fabrication process development and systematic device design. The proposed $\kappa$-taper achieves around 75\% improvement in transition efficiency and misalignment tolerance compared to L-taper, providing unprecedented design flexibility for dense 3D photonic integrated circuits. In the dual-layer SiN demonstration, the inter-layer transition loss maintains as low as 5.56 mdB/coupler with excellent wafer-scale uniformity. The proposed method establishes a clear and practical road-map for reliable, high‑density 3D PICs mass manufacturing, demonstrating the optimal design, fabrication process control, and yield needed to translate 3D photonics from prototype to production. It also provides a dependable platform for high‑dimensional combination of distinct photonic materials, enabling seamless integration of passive waveguides, active elements and nonlinear media across stacked layers for intriguing photonic applications.

\bibliography{main.bib}

@article{jin2021hertz,
  title={Hertz-linewidth semiconductor lasers using CMOS-ready ultra-high-Q microresonators},
  author={Jin, Warren and Yang, Qi-Fan and Chang, Lin and Shen, Boqiang and Wang, Heming and Leal, Mark A and Wu, Lue and Gao, Maodong and Feshali, Avi and Paniccia, Mario and others},
  journal={Nature Photonics},
  volume={15},
  number={5},
  pages={346--353},
  year={2021},
  publisher={Nature Publishing Group UK London}
}

@article{xu2019sin,
  title={SiN x--Si interlayer coupler using a gradient index metamaterial},
  author={Xu, Pengfei and Zhang, Yanfeng and Zhang, Shuailong and Chen, Yujie and Yu, Siyuan},
  journal={Optics letters},
  volume={44},
  number={5},
  pages={1230--1233},
  year={2019},
  publisher={Optical Society of America}
}

@article{jones2013ultra,
  title={Ultra-low crosstalk, CMOS compatible waveguide crossings for densely integrated photonic interconnection networks},
  author={Jones, Adam M and DeRose, Christopher T and Lentine, Anthony L and Trotter, Douglas C and Starbuck, Andrew L and Norwood, Robert A},
  journal={Optics express},
  volume={21},
  number={10},
  pages={12002--12013},
  year={2013},
  publisher={Optical Society of America}
}

@article{zhu2024low,
  title={Low-loss and polarization insensitive 32$\times$ 4 optical switch for ROADM applications},
  author={Zhu, Xiaotian and Wang, Xiang and Huang, Yanlu and Wu, Liyan and Zhao, Chunfei and Xiao, Mingzhu and Wang, Luyi and Davidson, Roy and Ou, Yanni and Little, Brent E and others},
  journal={Light: Science \& Applications},
  volume={13},
  number={1},
  pages={94},
  year={2024},
  publisher={Nature Publishing Group UK London}
}

@article{niels2026high,
  title={A high-speed heterogeneous lithium tantalate silicon photonics platform},
  author={Niels, Margot and Vanackere, Tom and Vissers, Ewoud and Zhai, Tingting and Nenezic, Patrick and Declercq, Jakob and Bruynsteen, C{\'e}dric and Niu, Shengpu and Moerman, Arno and Caytan, Olivier and others},
  journal={Nature Photonics},
  pages={1--7},
  year={2026},
  publisher={Nature Publishing Group UK London}
}

@article{zhang2018high,
  title={High-density wafer-scale 3-D silicon-photonic integrated circuits},
  author={Zhang, Yu and Ling, Yi-Chun and Zhang, Yichi and Shang, Kuanping and Yoo, SJ Ben},
  journal={IEEE Journal of Selected Topics in Quantum Electronics},
  volume={24},
  number={6},
  pages={1--10},
  year={2018},
  publisher={IEEE}
}

@article{kang2014amorphous,
  title={Amorphous-silicon inter-layer grating couplers with metal mirrors toward 3-D interconnection},
  author={Kang, JoonHyun and Atsumi, Yuki and Hayashi, Yusuke and Suzuki, Junichi and Kuno, Yuki and Amemiya, Tomohiro and Nishiyama, Nobuhiko and Arai, Shigehisa},
  journal={IEEE Journal of Selected Topics in Quantum Electronics},
  volume={20},
  number={4},
  pages={317--322},
  year={2014},
  publisher={IEEE}
}

@article{sodagar2014high,
  title={High-efficiency and wideband interlayer grating couplers in multilayer Si/SiO 2/SiN platform for 3D integration of optical functionalities},
  author={Sodagar, Majid and Pourabolghasem, Reza and Eftekhar, Ali A and Adibi, Ali},
  journal={Optics express},
  volume={22},
  number={14},
  pages={16767--16777},
  year={2014},
  publisher={Optica Publishing Group}
}

@article{wan2016grating,
  title={Grating design for interlayer optical interconnection of in-plane waveguides},
  author={Wan, Congshan and Gaylord, Thomas K and Bakir, Muhannad S},
  journal={Applied Optics},
  volume={55},
  number={10},
  pages={2601--2610},
  year={2016},
  publisher={Optica Publishing Group}
}

@article{sacher2017tri,
  title={Tri-layer silicon nitride-on-silicon photonic platform for ultra-low-loss crossings and interlayer transitions},
  author={Sacher, Wesley D and Mikkelsen, Jared C and Dumais, Patrick and Jiang, Jia and Goodwill, Dominic and Luo, Xianshu and Huang, Ying and Yang, Yisu and Bois, Antoine and Lo, Patrick Guo-Qiang and others},
  journal={Optics express},
  volume={25},
  number={25},
  pages={30862--30875},
  year={2017},
  publisher={Optical Society of America}
}

@article{yao2022bridging,
  title={Bridging the gap between resonance and adiabaticity: a compact and highly tolerant vertical coupling structure},
  author={Yao, Chunhui and Cheng, Qixiang and Roelkens, G{\"u}nther and Penty, Richard},
  journal={Photonics Research},
  volume={10},
  number={9},
  pages={2081--2090},
  year={2022},
  publisher={Chinese Laser Press and Optica Publishing Group}
}

@article{taras2021shortcuts,
  title={Shortcuts to adiabaticity in waveguide couplers--theory and implementation},
  author={Taras, Adam K and Tuniz, Alessandro and Bajwa, Musawer A and Ng, Vincent and Dawes, Judith M and Poulton, Christopher G and De Sterke, C Martijn},
  journal={Advances in Physics: X},
  volume={6},
  number={1},
  pages={1894978},
  year={2021},
  publisher={Taylor \& Francis}
}

@article{zhang20253d,
  title={3D Heterogeneous Integration of Silicon Nitride and Aluminum Nitride on Sapphire toward Ultrawideband Photonics Integrated Circuits},
  author={Zhang, Liang and Guo, Yanan and Wang, Junxi and Li, Jinmin and Yan, Jianchang},
  journal={ACS Photonics},
  volume={12},
  number={5},
  pages={2538--2547},
  year={2025},
  publisher={ACS Publications}
}

@article{shang2015low,
  title={Low-loss compact multilayer silicon nitride platform for 3D photonic integrated circuits},
  author={Shang, Kuanping and Pathak, Shibnath and Guan, Binbin and Liu, Guangyao and Yoo, SJB},
  journal={Optics Express},
  volume={23},
  number={16},
  pages={21334--21342},
  year={2015},
  publisher={Optical Society of America}
}

@article{van2024methodical,
  title={A methodical approach to design adiabatic waveguide couplers for heterogeneous integrated photonics},
  author={Van Asch, Jef and Kandeel, Ahmed and He, Junwen and Missinne, Jeroen and Bienstman, Peter and Van Thourhout, Dries and Van Steenberge, Geert and Van Campenhout, Joris},
  journal={Journal of Physics: Photonics},
  volume={6},
  number={4},
  pages={045013},
  year={2024},
  publisher={IOP Publishing}
}

@article{zhang2020scal,
  title={Scalable 3D silicon photonic electronic integrated circuits and their applications},
  author={Zhang, Yu and Samanta, Anirban and Shang, Kuanping and Yoo, SJ Ben},
  journal={IEEE Journal of Selected Topics in Quantum Electronics},
  volume={26},
  number={2},
  pages={1--10},
  year={2020},
  publisher={IEEE}
}

@inproceedings{Xia2025ThreeD,
  author={Xia, Yu and Huang, Yuhao and Liu, Mingfei and Wang, Jie and Li, Zheng and Li, Yuemin and Fu, Yunqi and Xiang, Chao},
  booktitle={2025 European Conference on Optical Communications (ECOC)}, 
  title={3D Silicon Nitride Waveguide Interposers for High-Density Scale-up Chiplet Interconnects}, 
  year={2025},
  volume={},
  number={},
  pages={1-4}
}

@article{kaur2021hybrid,
  title={Hybrid and heterogeneous photonic integration},
  author={Kaur, Paramjeet and Boes, Andreas and Ren, Guanghui and Nguyen, Thach G and Roelkens, Gunther and Mitchell, Arnan},
  journal={APL photonics},
  volume={6},
  number={6},
  year={2021},
  publisher={AIP Publishing}
}

@article{sacher2018mono,
  title={Monolithically integrated multilayer silicon nitride-on-silicon waveguide platforms for 3-D photonic circuits and devices},
  author={Sacher, Wesley D and Mikkelsen, Jared C and Huang, Ying and Mak, Jason CC and Yong, Zheng and Luo, Xianshu and Li, Yu and Dumais, Patrick and Jiang, Jia and Goodwill, Dominic and others},
  journal={Proceedings of the IEEE},
  volume={106},
  number={12},
  pages={2232--2245},
  year={2018},
  publisher={IEEE}
}

@article{Margalit2021Per,
    author = {Margalit, Near and Xiang, Chao and Bowers, Steven M. and Bjorlin, Alexis and Blum, Robert and Bowers, John E.},
    title = {Perspective on the future of silicon photonics and electronics},
    journal = {Applied Physics Letters},
    volume = {118},
    number = {22},
    pages = {220501},
    year = {2021},
    month = {06}
}

@article{tran2022ext,
  title={Extending the spectrum of fully integrated photonics to submicrometre wavelengths},
  author={Tran, Minh A and Zhang, Chong and Morin, Theodore J and Chang, Lin and Barik, Sabyasachi and Yuan, Zhiquan and Lee, Woonghee and Kim, Glenn and Malik, Aditya and Zhang, Zeyu and others},
  journal={Nature},
  volume={610},
  number={7930},
  pages={54--60},
  year={2022},
  publisher={Nature Publishing Group UK London}
}

@article{churaev2023hg,
  title={A heterogeneously integrated lithium niobate-on-silicon nitride photonic platform},
  author={Churaev, Mikhail and Wang, Rui Ning and Riedhauser, Annina and Snigirev, Viacheslav and Bl{\'e}sin, Terence and M{\"o}hl, Charles and Anderson, Miles H and Siddharth, Anat and Popoff, Youri and Drechsler, Ute and others},
  journal={Nature communications},
  volume={14},
  number={1},
  pages={3499},
  year={2023},
  publisher={Nature Publishing Group UK London}
}

@article{xiang20233d,
  title={3D integration enables ultralow-noise isolator-free lasers in silicon photonics},
  author={Xiang, Chao and Jin, Warren and Terra, Osama and Dong, Bozhang and Wang, Heming and Wu, Lue and Guo, Joel and Morin, Theodore J and Hughes, Eamonn and Peters, Jonathan and others},
  journal={Nature},
  volume={620},
  number={7972},
  pages={78--85},
  year={2023},
  publisher={Nature Publishing Group UK London}
}

@article{ji2024efficient,
  title={Efficient mass manufacturing of high-density, ultra-low-loss Si3N4 photonic integrated circuits},
  author={Ji, Xinru and Ning Wang, Rui and Liu, Yang and Riemensberger, Johann and Qiu, Zheru and Kippenberg, Tobias J},
  journal={Optica},
  volume={11},
  number={10},
  pages={1397--1407},
  year={2024},
  publisher={Optica Publishing Group}
}

@article{liu2021high,
  title={High-yield, wafer-scale fabrication of ultralow-loss, dispersion-engineered silicon nitride photonic circuits},
  author={Liu, Junqiu and Huang, Guanhao and Wang, Rui Ning and He, Jijun and Raja, Arslan S and Liu, Tianyi and Engelsen, Nils J and Kippenberg, Tobias J},
  journal={Nature communications},
  volume={12},
  number={1},
  pages={2236},
  year={2021},
  publisher={Nature Publishing Group UK London}
}

@inproceedings{bauters2011ultra,
  title={Ultra-low-loss single-mode Si3N4 waveguides with 0.7 dB/m propagation loss},
  author={Bauters, Jared F and Heck, Martijn JR and John, Demis D and Tien, Ming-Chun and Li, Wenzao and Barton, Jon S and Blumenthal, Daniel J and Bowers, John E and Leinse, Arne and Heideman, Ren{\'e} G},
  booktitle={European conference and exposition on optical communications},
  pages={Th--12},
  year={2011},
  organization={Optica Publishing Group}
}

@article{pfeiffer2016photonic,
  title={Photonic Damascene process for integrated high-Q microresonator based nonlinear photonics},
  author={Pfeiffer, Martin Hubert Peter and Kordts, Arne and Brasch, Victor and Zervas, Michael and Geiselmann, Michael and Jost, John D and Kippenberg, Tobias J},
  journal={Optica},
  volume={3},
  number={1},
  pages={20--25},
  year={2016},
  publisher={Optical Society of America}
}

@article{pfeiffer2018photonic,
  title={Photonic damascene process for low-loss, high-confinement silicon nitride waveguides},
  author={Pfeiffer, Martin Hubert Peter and Herkommer, Clemens and Liu, Junqiu and Morais, Tiago and Zervas, Michael and Geiselmann, Michael and Kippenberg, Tobias J},
  journal={IEEE Journal of selected topics in quantum electronics},
  volume={24},
  number={4},
  pages={1--11},
  year={2018},
  publisher={IEEE}
}

@article{wu2020stress,
  title={Stress-released Si3N4 fabrication process for dispersion-engineered integrated silicon photonics},
  author={Wu, Kaiyi and Poon, Andrew W},
  journal={Optics Express},
  volume={28},
  number={12},
  pages={17708--17722},
  year={2020},
  publisher={Optical Society of America}
}

@article{liu2025fabrication,
  title={Fabrication of Ultra-Low-Loss, Dispersion-Engineered Silicon Nitride Photonic Integrated Circuits via Silicon Hardmask Etching},
  author={Liu, Shuai and Zhang, Yuheng and Hariri, Abdulkarim and Al-Hallak, Abdur-Raheem and Zhang, Zheshen},
  journal={ACS Photonics},
  volume={12},
  number={2},
  pages={1039--1046},
  year={2025},
  publisher={ACS Publications}
}

@article{chiles2018de,
  title={Deuterated silicon nitride photonic devices for broadband optical frequency comb generation},
  author={Chiles, Jeff and Nader, Nima and Hickstein, Daniel D and Yu, Su Peng and Briles, Travis Crain and Carlson, David and Jung, Hojoong and Shainline, Jeffrey M and Diddams, Scott and Papp, Scott B and others},
  journal={Optics Letters},
  volume={43},
  number={7},
  pages={1527--1530},
  year={2018},
  publisher={Optical Society of America}
}

@article{zhang2024low,
  title={Low-Temperature Sputtered Ultralow-Loss Silicon Nitride for Hybrid Photonic Integration},
  author={Zhang, Shuangyou and Bi, Toby and Harder, Irina and Ohletz, Olga and Gannott, Florentina and Gumann, Alexander and Butzen, Eduard and Zhang, Yaojing and Del'Haye, Pascal},
  journal={Laser \& Photonics Reviews},
  volume={18},
  number={4},
  pages={2300642},
  year={2024},
  publisher={Wiley Online Library}
}

@article{bose2024anneal,
  title={Anneal-free ultra-low loss silicon nitride integrated photonics},
  author={Bose, Debapam and Harrington, Mark W and Isichenko, Andrei and Liu, Kaikai and Wang, Jiawei and Chauhan, Nitesh and Newman, Zachary L and Blumenthal, Daniel J},
  journal={Light: Science \& Applications},
  volume={13},
  number={1},
  pages={156},
  year={2024},
  publisher={Nature Publishing Group UK London}
}

@article{liang2021fully,
  title={A fully numerical method for designing efficient adiabatic mode evolution structures (adiabatic taper, coupler, splitter, mode converter) applicable to complex geometries},
  author={Liang, Tu-Lu and Tu, Yongming and Chen, Xi and Huang, Yingyan and Bai, Qiang and Zhao, Yaying and Zhang, Junchi and Yuan, Yutong and Li, Junyu and Yi, Fei and others},
  journal={Journal of Lightwave Technology},
  volume={39},
  number={17},
  pages={5531--5547},
  year={2021},
  publisher={IEEE}
}

@article{ji2022compact,
  title={Compact, spatial-mode-interaction-free, ultralow-loss, nonlinear photonic integrated circuits},
  author={Ji, Xinru and Liu, Junqiu and He, Jijun and Wang, Rui Ning and Qiu, Zheru and Riemensberger, Johann and Kippenberg, Tobias J},
  journal={Communications Physics},
  volume={5},
  number={1},
  pages={84},
  year={2022},
  publisher={Nature Publishing Group UK London}
}


\medskip

\vspace{3 mm}

\section*{Methods}

\subsection*{Comparison of SiN deposition method and fabrication process flow}
To release stress and reduce the wafer bow, process flow based on low-pressure chemical vapor deposition (LPCVD) SiN usually demands complicated processes, such as the photonic Damascene process~\cite{pfeiffer2016photonic}, the extra backside SiN removal process~\cite{liu2021high, pfeiffer2018photonic}, the use of nonstandard 4-inch wafers with 700 $\mu$m-thick substrate~\cite{pfeiffer2018photonic}, or multi-step deposition process assisted by exquisitely designed patterns~\cite{wu2020stress, liu2025fabrication}. The number of these process steps multiplies as SiN layer increases, further complicating the process flow and limiting process reproducibility. On the other hand, high-temperature annealing ($\sim$1150$^{\circ}$C) is crucial and indispensable for obtaining high-quality SiN, but it also leads to considerable stress regardless of LPCVD ($\sim$780$^{\circ}$C) or PECVD ($<$400$^{\circ}$C). Although preparing SiN by sputtering \cite{zhang2024low} and deuterated silane \cite{chiles2018de, bose2024anneal} can avoid high-temperature process, the promotion of these schemes is largely constrained by the particularities of the equipment and sources employed. 

On this basis, SiN prepared by PECVD, which intrinsically possesses low stress, provides convenient stress management and fabrication repeatability without compromising the optical propagation loss after high-temperature annealing. The next task is to tackle the stress resulting from densification during the annealing of PECVD SiN. 

Here, we etch SRT prior to each annealing cycle to divide the continuous SiN film into localized segments to release the stress caused by annealing. The detailed fabrication process flow is:

Step 1: Wafer preparation. 4 $\mu$m-thick oxide layer through dry-wet oxidation is formed on standard 4-inch silicon wafers as the WG lower cladding. Wafers with wafer bow below 5 $\mu$m are chosen and then patterned with stepper alignment marks.

Step 2: SiN deposition. PECVD SiN is deposited onto the prepared wafer at 360$^{\circ}$C with the plasma switching between a high (13.56 MHz) and a low frequency (100 kHz). The power settings and duration for each frequency mode are optimized to achieve low film stress, with 4 measured samples demonstrating an average stress value of 51.18 MPa (See SI Section II). The film thickness should be slightly increased so that the total thickness conforms to the design value after annealing (See SI Section I, film thickness reduction of 85\%). In this work, a 470 nm PECVD SiN film should be deposited for a designed 400-nm-thick SiN WG.

Step 3: SRT formation and annealing. The deposited SiN film is fully etched with SRT that is designed to avoid WG areas and prevent large blank regions (including the area outside the exposure zone of stepper lithography, see SI Section I). The wafer then undergoes 1150$^{\circ}$C annealing in N$_2$ atmosphere for 2 hours.

Step 4: WG patterning. An anti-reflective layer and photo resist are sequentially spin-coated onto the wafer. After lithography and development, the photo resist is reflowed to reduce sidewall roughness. Then, the patterns are transferred to SiN using fluorine-based dry etch. 

Step 5: SiO$_2$ cladding. Similar to SiN deposition, PECVD TEOS-SiO$_2$ is deposited onto the wafer with optimized plasma settings to lower the stress. SRT formation and annealing are performed to SiO$_2$ again for stress and loss control. Before depositing the second layer of SiN, SiO$_2$ cladding is etched back and planarized.

\subsection*{Design of $\kappa$-engineered taper}
The linear taper is divided into 20 segments (see SI Section V for details of parameters optimization), and then three points are fixed: the phase-matched point, the start point, and the end point. Due to the identical material of both tapers, $\kappa$ exhibits symmetry about the phase-matched point, resulting in derivatives at the start and end points being opposite in sign. To obtain the gentle curve $H(x)$ by Hermite interpolation, the $1^{st}$ derivative at the phase-matched point is set to zero for a smoothly varying $\kappa$, and the only parameter to be determined is the $1^{st}$ derivative at the start (end) point, denoted as $h$. By roughly sweeping several values of $h$ and observing the taper transmission versus length, one can select an $h$ that yields a smooth transmission line. Then, $\kappa$ as well as the taper widths can be remapped onto the smooth curve $H(x)$. During this $\kappa$ remapping and taper geometry reshaping process, the taper length is set to a normalized value for convenience. In the numerical simulation or layout design of $\kappa$-taper, once the total length is determined, the length of each segment can be scaled proportionally.

\subsection*{Device characterization}

In the experimental setup, light emitted from a Keysight N7778C tunable laser is first passed through a polarization controller to obtain transverse electric (TE) modes and then edge-coupled into the chip using a lensed fiber. The transmitted output from the chip is subsequently edge-coupled into a second lensed fiber, routed through an attenuator, and finally directed to a NewPort 1811 photodetector (PD) for detection. The resulting photocurrent signal is sampled and recorded by a Keysight MSOX6004A oscilloscope. The system is configured to perform a wavelength sweep from 1480 nm to 1620 nm at a speed of 80 nm/s. Throughout this sweep, the optical power received by the PD is maintained near -18 dBm to acquire an optimal signal-to-noise ratio as well as prevent damage to PD. This configuration features a wavelength step of $5.12\times10^{-6}$ nm, which is fine enough for resonance dip measurement.

\vspace{3 mm}

\noindent \textbf{Data Availability}
The data that supports the plots within this paper and other findings of this study are available from the corresponding author upon reasonable request.
\vspace{1 mm}

\noindent \textbf{Code Availability}
The codes that support findings of this study are available from the corresponding author upon reasonable request.
\vspace{1 mm}

\noindent \textbf{Acknowledgments}
We thank the funding support from the National Key R\&D Program of China (2024YFA1409300), the Research Grants Council of Hong Kong (C7143-25Y, N\_HKU774\_25, T46-705/23-R, STG3/E-704/23-N, STG3/E-104/25-N), the Innovation and Technology Commission of Hong Kong (GHP/230/22GD), the National Natural Science Foundation of China (6232290014), the Guangdong Provincial Quantum Science Strategic Initiative (GDZX2304004, GDZX2404002), and the Croucher Foundation. 
We thank Mr Henry Chun Fai Yeung, Mr Ho Li, and other technicians in Nanosystem Fabrication Facility (CWB) of HKUST for help in device fabrication.
We thank Dr. Ruixuan Chen for the help in figure preparation.
\vspace{1 mm}

\noindent\textbf{Author Contributions} 
Concepts were conceived by Y.H. and C.X.. 
Structures were designed by Y.H..
Samples were fabricated by Y.H. with the assistance of Y.F., Z.L., Y.H. and Z.G..
Measurements were performed by Y.H., Y.F., Y.X., Y.L. and M.L..
All authors contributed to the writing of the manuscript. 
The project was supervised by C.X..
\vspace{1 mm}

\noindent \textbf{Competing Interests} The authors declare no competing interests.

\vspace{1 mm}

\noindent \textbf{Author Information} Correspondence and requests for materials should be addressed to C.X. (cxiang@eee.hku.hk).

\end{document}